\documentclass[a4paper,11pt]{article}
\usepackage{pos}

\title{Localisation of Dirac modes in finite-temperature
$\mathbb{Z}_2$ gauge theory on the lattice}
\author*[a]{Gy\"orgy Baranka}
\author[a]{Matteo Giordano}

\affiliation[a]{ELTE E\"otv\"os Lor\'and University, Institute for Theoretical Physics,\\
 P\'azm\'any P\'eter s\'et\'any 1/A, H-1117, Budapest, Hungary}

\emailAdd{barankagy@caesar.elte.hu}
\emailAdd{giordano@bodri.elte.hu}

\abstract{The low-lying Dirac modes become localised at the finite-temperature
transition in QCD and in other gauge theories, suggesting a
general connection between their localisation and deconfinement. The
simplest model where this connection can be tested is $\mathbb{Z}_2$ gauge
theory in 2+1 dimensions. We show that in this model the low modes in the staggered Dirac spectrum are delocalised in the confined phase and become localised in the deconfined phase. We also show that localised modes correlate with disorder in the
Polyakov loop configuration, in agreement with the ``sea/island''
picture of localisation, and with negative plaquettes. These results further support the conjecture
that localisation and deconfinement are closely related.}

\FullConference{%
 The 38th International Symposium on Lattice Field Theory, LATTICE2021
  26th-30th July, 2021
  Zoom/Gather@Massachusetts Institute of Technology
}

\begin{document}
\maketitle

\section{Introduction}
In recent years it has become apparent that there is a strong relation between the confining properties of gauge theories and the localisation properties of Dirac eigenmodes. It is now established that the low modes of the Dirac operator become localised in the high temperature phase of QCD, up to a temperature-dependent point in the spectrum \cite{2007,2012,2016,holicki2018anderson}. This phenomenon is found in other gauge theories and gauge related models as well \cite{2001,2020,2021,PhysRevD.95.074503,PhysRevD.97.014502,PhysRevLett.104.031601,PhysRevLett.105.192001}. For a recent review see \cite{summurize}. This situation was put into a relation with the ordering of Polyakov loops in the high temperature phase in the so-called ``sea/islands'' picture of localisation, according to which the low Dirac modes localise near ``energetically'' favourable fluctuations in the ordered sea of Polyakov loops. Clear numerical evidence for this correlation was found \cite{PhysRevD.84.034505,2016,holicki2018anderson}. 

Localisation of eigenmodes caused by disorder is a well-studied phenomenon in condensed matter physics~\cite{condensed}. Detailed numerical studies have confirmed that localisation of the low-lying Dirac modes in QCD at high temperature shares the same critical features with three-dimensional Hamiltonians with on-site disorder in the appropriate symmetry class \cite{2014,Nishigaki:2013uya,33ofpaper}. This is compatible with the conjectured connection between localised modes and Polyakov loop fluctuations, that have the right properties to be the relevant source of disorder.

An interesting aspect of the sea/islands picture is its simplicity, as it basically only requires ordering of the Polyakov loops. This suggests that localised low Dirac modes will generally be found in the deconfined phase of a gauge theory. This has been confirmed by numerical studies in a variety of models  \cite{PhysRevD.97.014502,PhysRevLett.104.031601,PhysRevLett.105.192001,2001,2020,2021,PhysRevD.95.074503,17,19}. All these findings strongly support the connection between localisation and deconfinement. In Ref. \cite{PhysRevD.104.054513} we pushed this connection to the limit and studied the simplest gauge group where a deconfining transition is found, i.e., $\mathbb{Z}_2$ gauge theory in 2+1 dimensions. 

\section{$\mathbb{Z}_2$ gauge theory}

The phase diagram of $\mathbb{Z}_2$ gauge theory with standard Wilson action on a cubic $N_t \times N_s^2$ lattice was studied in detail in Ref. \cite{CASELLE1996435}, exploiting the duality with the three dimensional Ising model.
A second order phase transition was found at a critical $\beta_c=\beta_c(N_t)$ separating the confined ($\beta < \beta_c$) and deconfined ($\beta > \beta_c$) phases. The  relevant order parameter of the transition is the expectation value of the Polyakov loop,
\begin{equation}
    P(\Vec{x}) = \prod_{z=0}^{N_t-1} U_1(t,\Vec{x}).
\end{equation}
In the thermodinamic limit, this quantity vanishes in the confined phase, while it is nonzero in the deconfined phase, where $P(\Vec{x})$ takes mostly either the $+1$ or the $-1$ value. The physically relevant case, corresponding to the presence of infinitely heavy fermions, is when Polyakov loops prefer the $+1$ value. We call this the ``physical'' sector, while the other case is the ``unphysical'' sector. 

In Ref.~\cite{PhysRevD.104.054513} we studied the staggered Dirac operator in the background of $\mathbb{Z}_2$ gauge field configurations. For $\mathbb{Z}_2$ gauge group the staggered operator reads
\begin{equation}
    D^{\mathrm{stag}}_{n,n'} = \frac{1}{2} \sum_{\mu=1}^3 \eta_\mu(n)(U_\mu(n) \delta_{n+\hat{\mu},n'}-U_\mu(n-\hat{\mu}) \delta_{n-\hat{\mu},n'}), \: \; \; \eta_\mu(n) = (-1)^{\sum_{\nu<\mu}n_\nu},
\end{equation}
where $U_\mu(n)= \pm 1$, with periodic boundary conditions in the spatial directions and antiperiodic boundary conditions in the temporal direction. The staggered operator is anti-Hermitian and so has purely imaginary eigenvalues,
\begin{equation}
    D^{\textrm{stag}} \psi_l =  i \lambda_l \psi_l, \: \; \lambda_l \in \mathbb{R}.
\end{equation}
The chiral property $\{ \varepsilon, D^{\textrm{stag}} \} = 0$, with $\varepsilon(n) = (-1)^{\sum_{\mu=1}^3 n_\mu}$ implies furthermore 
\begin{equation}
\label{sym}
    D^{\textrm{stag}} \varepsilon \psi_l = -i \lambda_l \varepsilon \psi_l,
\end{equation}
so that the spectrum is symmetric about zero.

\section{Localisation of Dirac eigenmodes}
The simplest way to detect localised modes is by using the so-called participation ratio,
\begin{equation}
    \textrm{PR}_l =  \frac{1}{N_t V} \Big( \sum_n |\psi_l(n)|^4\Big)^{-1} ,
\end{equation}
which measures the fraction of spacetime that the $l^{\textrm{th}}$ mode effectively occupies. As $V \xrightarrow{} \infty$, this quantity tends to zero for localised modes, while it tends to a finite value for fully delocalised modes with $|\psi_l(n)|^2 \sim 1/(V N_t)$.  

To identify the spectral regions where modes are localised, one divides the spectrum into small bins (ideally infinitesimal), and averages the PR over modes within each bin and over gauge configurations,
\begin{equation}
    \textrm{PR}(\lambda, N_s)= \frac{\langle \sum_l \delta(\lambda-\lambda_l)\textrm{PR}_l\rangle}{\sum_l \delta(\lambda-\lambda_l) \rangle},
\end{equation}
studying then the behaviour of $\textrm{PR}(\lambda,N_s)$ as $N_s$ is changed.  The same was done for the other observables that we used in this study. At large $N_s$ one expects the following behaviour of $\textrm{PR}(\lambda,N_s)$,
\begin{equation}
    \textrm{PR}(\lambda,N_s) \simeq c(\lambda) N_s^{\alpha(\lambda)-2},
\end{equation}
with some volume-independent $c(\lambda)$, where $\alpha(\lambda)$ is the fractal dimension of the eigenmodes in the given spectral region. For fully delocalised modes $\alpha = 2 $, while for localised modes $\alpha = 0$. The fractal dimension can be extracted from a pair of different system sizes ($N_{s1,s2}$),
\begin{equation}
    \alpha(\lambda) = 2 + \log \Bigg( \frac{\textrm{PR}(\lambda,N_{s1})}{\textrm{PR}(\lambda,N_{s2})} \Bigg) \Bigg/ \log \Bigg( \frac{N_{s1}}{N_{s2}} \Bigg),
\end{equation} 
for sufficiently large sizes.

It is expected from the sea/islands picture of localisation that localised modes strongly correlate with fluctuations of the Polyakov loops. To test this correlation one can introduce the following observable,
\begin{equation}
    \mathscr{P}(\lambda) = \frac{\langle \sum_l \delta(\lambda-\lambda_l)\sum_{t,\Vec{x}}P(\Vec{x})|\psi_l(t,\Vec{x})|^2\rangle}{\sum_l \delta(\lambda-\lambda_l) \rangle},
\end{equation}
i.e.,  the spatial average of Polyakov loops weighted by the fermion mode density, or the ``Polyakov loop seen by a mode'' \cite{PhysRevD.84.034505}. For fully delocalised modes one expects
\begin{equation}
    \sum_{t,\Vec{x}}P(\Vec{x})|\psi_l(t,\Vec{x})|^2 \simeq \frac{1}{N_t V} \sum_{t,\Vec{x}} P(\Vec{x}) = \frac{1}{V} \sum_{\Vec{x}} P(\Vec{x}) = \overline{P},
\end{equation}
and so approximately $\mathscr{P} \simeq \langle P \rangle$ in spectral regions where modes are extended. For localised modes (localised in a volume $V_0$) one expects instead
\begin{equation}
    \sum_{t,\Vec{x}}P(\Vec{x})|\psi_l(t,\Vec{x})|^2 \simeq \frac{1}{N_t V_0} \sum_{t,\Vec{x} \in V_0} P(\Vec{x}) = \frac{1}{V_0} \sum_{\Vec{x} \in V_0} P(\Vec{x}) = \overline{P}_{V_0},
\end{equation}
where $\overline{P}_{V_0}$ is the average Polyakov loop in the occupied volume.
In the deconfined phase in the physical sector, low modes are expected to prefer negative Polyakov loops, thus their $\mathscr{P}$ is expected to be much smaller than $\overline{P}$. 

It was pointed out in Ref.~\cite{negative} that the deconfinement transition is also characterised by the largest cluster of negative plaquettes ceasing to scale like the system size, so that it is not fully delocalised anymore in the deconfined phase. It is interesting then to check how localised modes and negative plaquettes correlate. To do this we have introduced the two following observables,
\begin{equation}
    \mathscr{U}(\lambda) = \frac{\langle \sum_l \delta(\lambda-\lambda_l)\sum_{n}A(n)|\psi_l(n)|^2\rangle}{\sum_l \delta(\lambda-\lambda_l) \rangle}, \: \;
     \Tilde{\mathscr{U}}(\lambda) = \frac{\langle \sum_l \delta(\lambda-\lambda_l)\sum_{n,A(n)>0}|\psi_l(n)|^2\rangle}{\sum_l \delta(\lambda-\lambda_l) \rangle},
\end{equation}
where $A(n)$ counts the number of negative plaquettes touching site $n$. $\mathscr{U}$ then measures the average number of negative plaquettes touched by the modes, while $\Tilde{\mathscr{U}}$ measures how much of the modes are touched by at least one negative plaquette. For delocalised modes one expects 
$
    \sum_n A(n) |\psi_l(n)|^2 \simeq 1/(N_t V) \sum A(n)$,
thus approximately $\mathscr{U} \simeq 6 (1-\langle U_{\mu \nu \rangle})$ in spectral regions where modes are extended. Estimating $\Tilde{\mathscr{U}}$ is instead not so straightforward, but $2(1-\langle U_{\mu \nu \rangle})$ turns out to provide an accurate lower bound.

\section{Numerical results}
We performed numerical simulations both in the confined and in the deconfined phase of the theory, on cubic lattices with $N_t=4$ and $N_s=20,24,28,32$. We used $\beta$ values on both sides of the critical coupling $\beta_c = 0.73107(2)$ \cite{CASELLE1996435}. Simulations were performed using the standard Metropolis algorithm. In this work we mainly show the results for the physical sector in the deconfined phase. Results for the unphysical sector can be found in \cite{PhysRevD.104.054513}.

For the free Dirac operator (i.e., on the trivial configuration $U_\mu(n) = 1, \forall n, \mu)$ in 2+1 dimensions the positive eigenvalues are in the region $\lambda_{(0)} \leq \lambda \leq \lambda_{(1)}$, where
$
    \lambda_{(0)} = 1/\sqrt{2}$ and $ \lambda_{(1)} = \sqrt{5/2}.
$
Based on these values the spectrum for nontrivial configurations can be divided into three regions: low modes ($\lambda < \lambda_{(0)}$), bulk modes ($\lambda_{(0)} \leq \lambda \leq \lambda_{(1)}$), and high modes ($\lambda > \lambda_{(1)}$).

\subsection{Participation ratio and fractal dimension}

\begin{figure}[t]
    \includegraphics[width=0.65\textwidth]{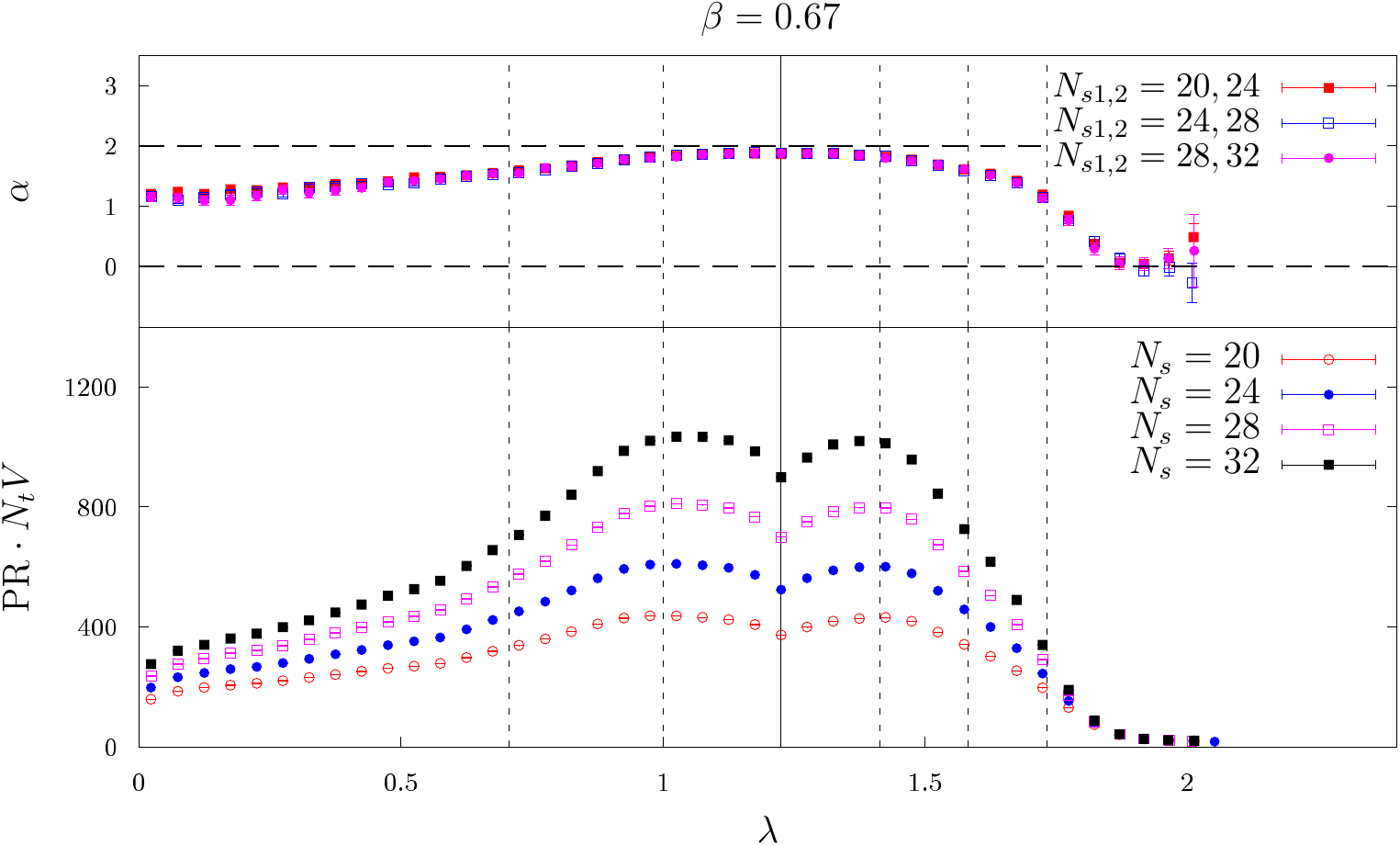}
    \centering
    \caption{Confined phase: mode size $\textrm{PR} \cdot N_t V$ (bottom panel)
    and fractal dimension $\alpha$ (top panel). }
    \label{fig:confpr}
\end{figure}

\begin{figure}[t]
    \includegraphics[width=0.7\textwidth]{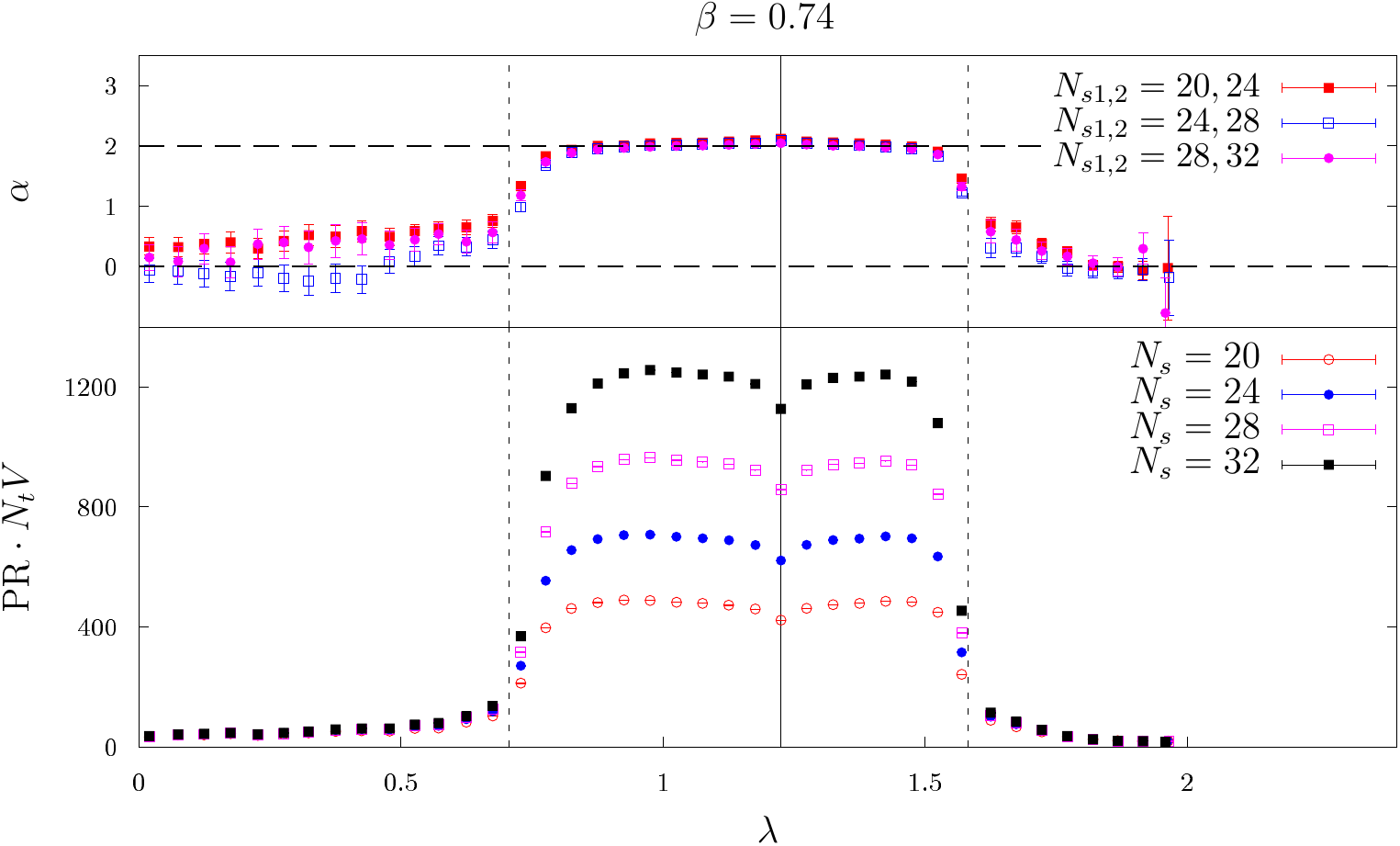}
    \centering
    \caption{Deconfined phase: mode size $\textrm{PR} \cdot N_t V $ (bottom panel)
and fractal dimension $\alpha$ (top panel). }
\label{fig:deconfpr}
\end{figure}
Our results for the ``size'' of the modes, $\textrm{PR} \cdot N_t V$, computed locally in the spectrum, are shown in Figs.~\ref{fig:confpr} and ~\ref{fig:deconfpr}. In the same figures we also show the fractal dimension $\alpha$. The statistical error was estimated by linear propagation of the jackknife errors on $\textrm{PR}(\lambda, N_s)$.

In the confined phase (Fig.~\ref{fig:confpr}) the size of the low modes increases with the volume, but not as fast as one would expect for fully delocalised modes. This means that low modes are delocalised but with a nontrivial fractal dimension, which starts from around $\alpha \approx 1$ for the lowest modes and increases towards 2 as one approaches the bulk. Bulk modes are delocalised with $\alpha$ close to 2. For high modes, the size of the occupied volume does not change with the system size. This means that these modes are localised, as shown by a fractal dimension compatible with zero. 

In the deconfined phase in the physical sector, the size of the low modes does not change with the volume and so the fractal dimension is zero. This indicates that these modes are localised. Bulk modes are fully delocalised ($\alpha\approx2$), and high modes are still localised.

\begin{figure}[t]
    \includegraphics[width=0.62\textwidth]{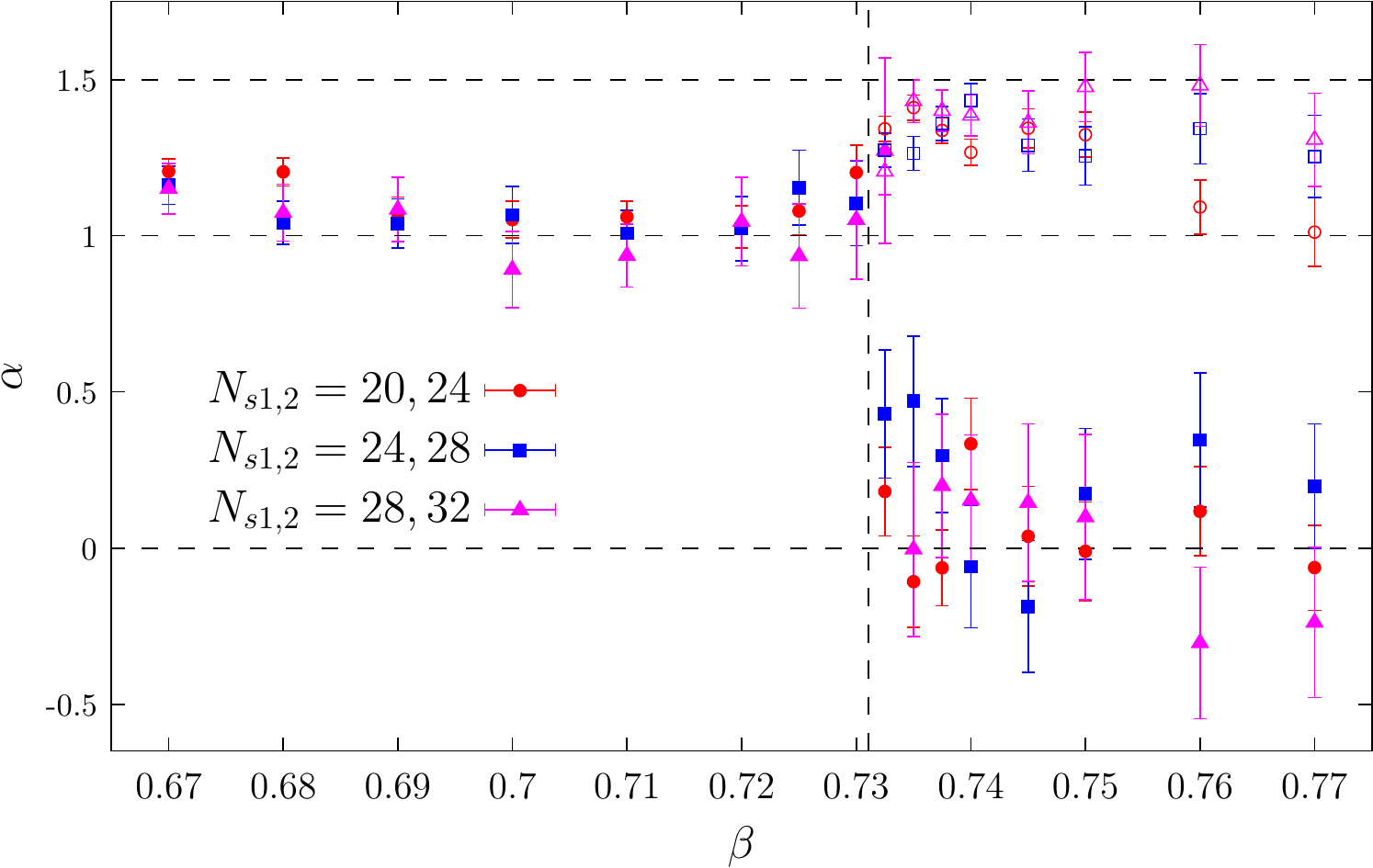}
    \centering
    \caption{Fractal dimension in the lowest spectral bin ($|\lambda|<0.05$), with physical and unphysical
sectors combined in the confined phase, and shown separately in the deconfined phase,
estimated with various pairs of system sizes. The critical coupling $\beta_c$ is marked by the
dashed vertical line.}
\label{fig:low}
\end{figure}

We also studied the fractal dimension of the near-zero modes (Fig. \ref{fig:low}) as a function of $\beta$. For $\beta<\beta_c$  the fractal dimension $\alpha$ is around 1, which means that these modes are delocalised with a nontrivial fractal dimension. For $\beta>\beta_c$ in the physical sector, near-zero modes get localised ($\alpha \approx 0$), while in the unphysical sector they do not and $\alpha > 1$.

\subsection{Polyakov loops}

\begin{figure}[t]
    \includegraphics[width=0.71\textwidth]{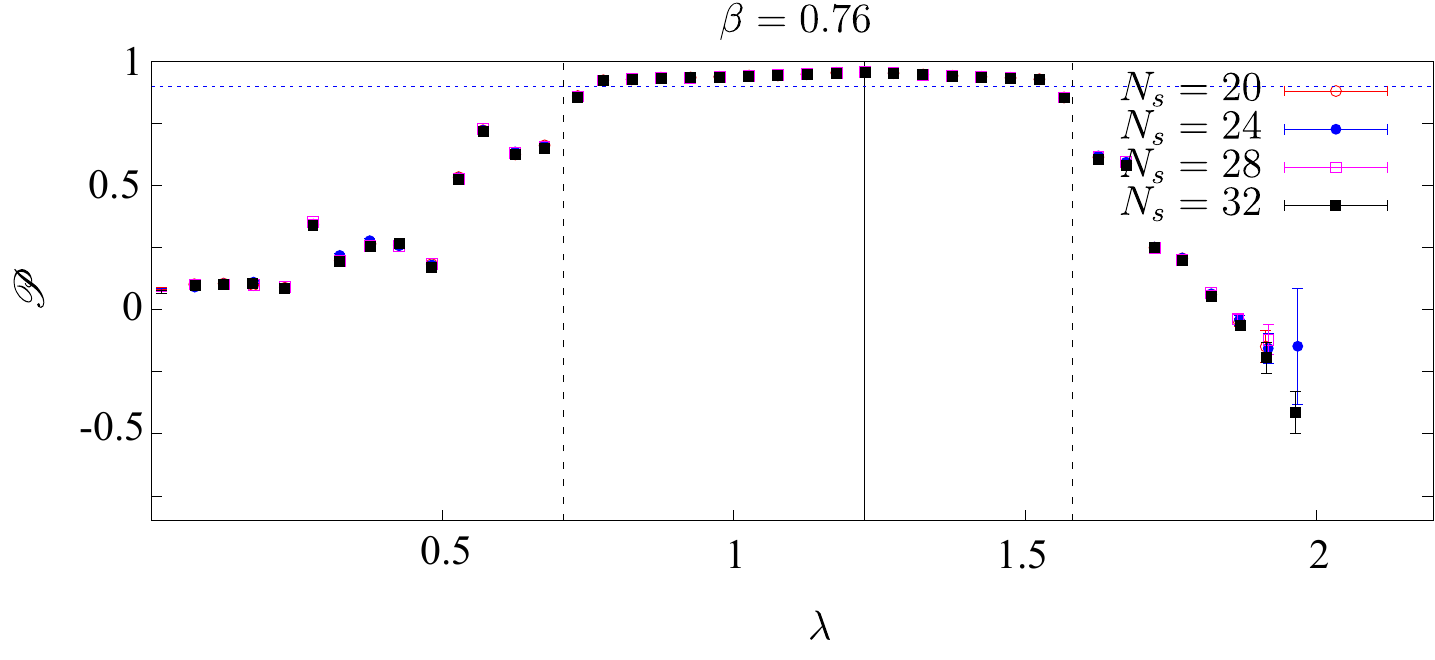}
    \centering
    \caption{Deconfined phase: Polyakov loop weighted by the modes (physical sector). The dashed blue line represents the average Polyakov loop.}
    \label{fig:poly}
\end{figure}

In Fig. \ref{fig:poly} we show the Polyakov loop weighted by the modes in the deconfined phase, physical sector. In spite of the fact that typically around 90\% of the Polyakov loops in a configuration are positive at this $\beta$, one finds $\mathscr{P} \approx 0$ for low modes, and $\mathscr{P} < 0$ for high modes. This shows that negative Polyakov loops and localised modes are strongly correlated, as expected.

\subsection{Negative plaquettes}

\begin{figure}[!h]
        \centering
        \includegraphics[width=0.5\linewidth]{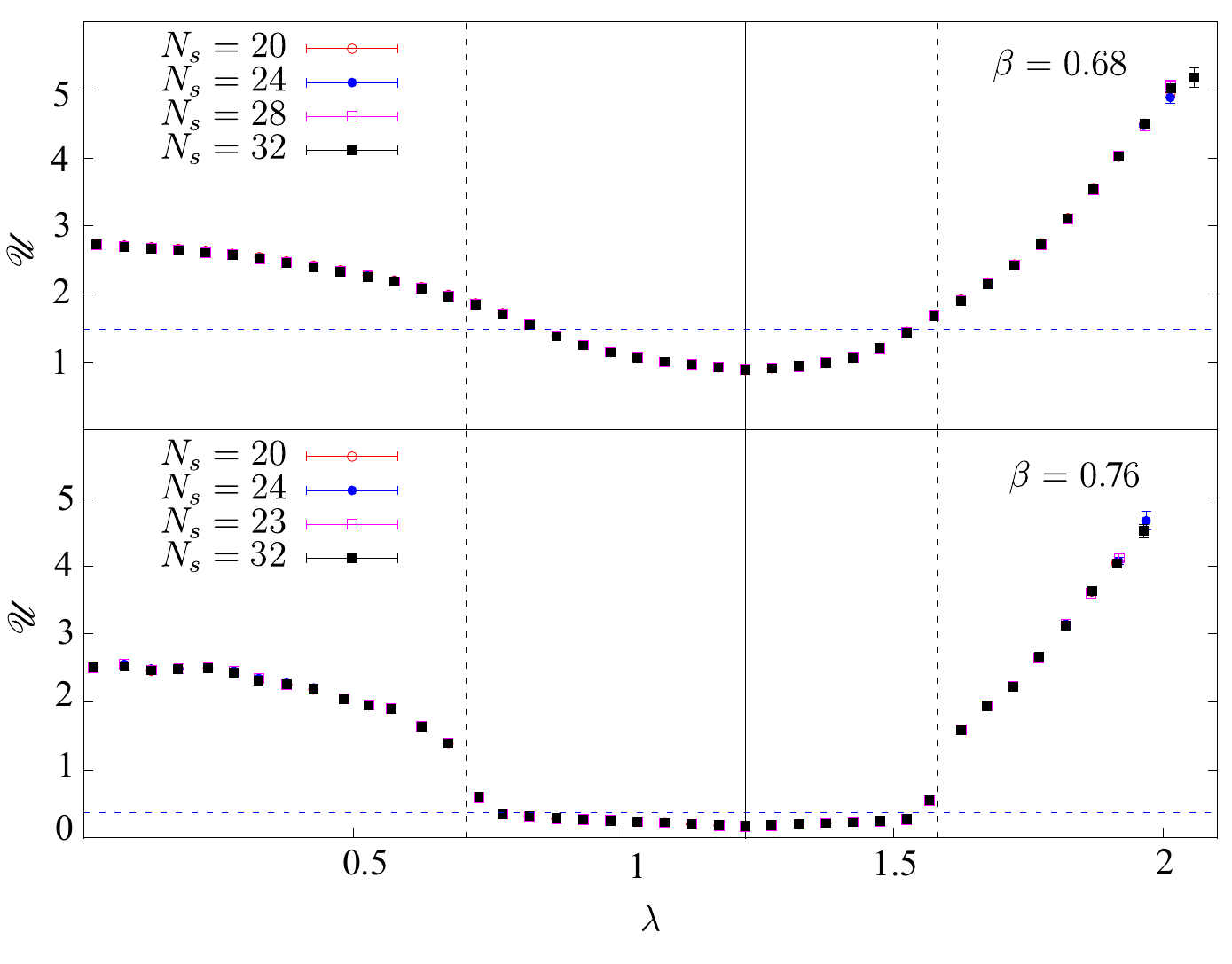}%
        \hfill
        \includegraphics[width=0.5\linewidth]{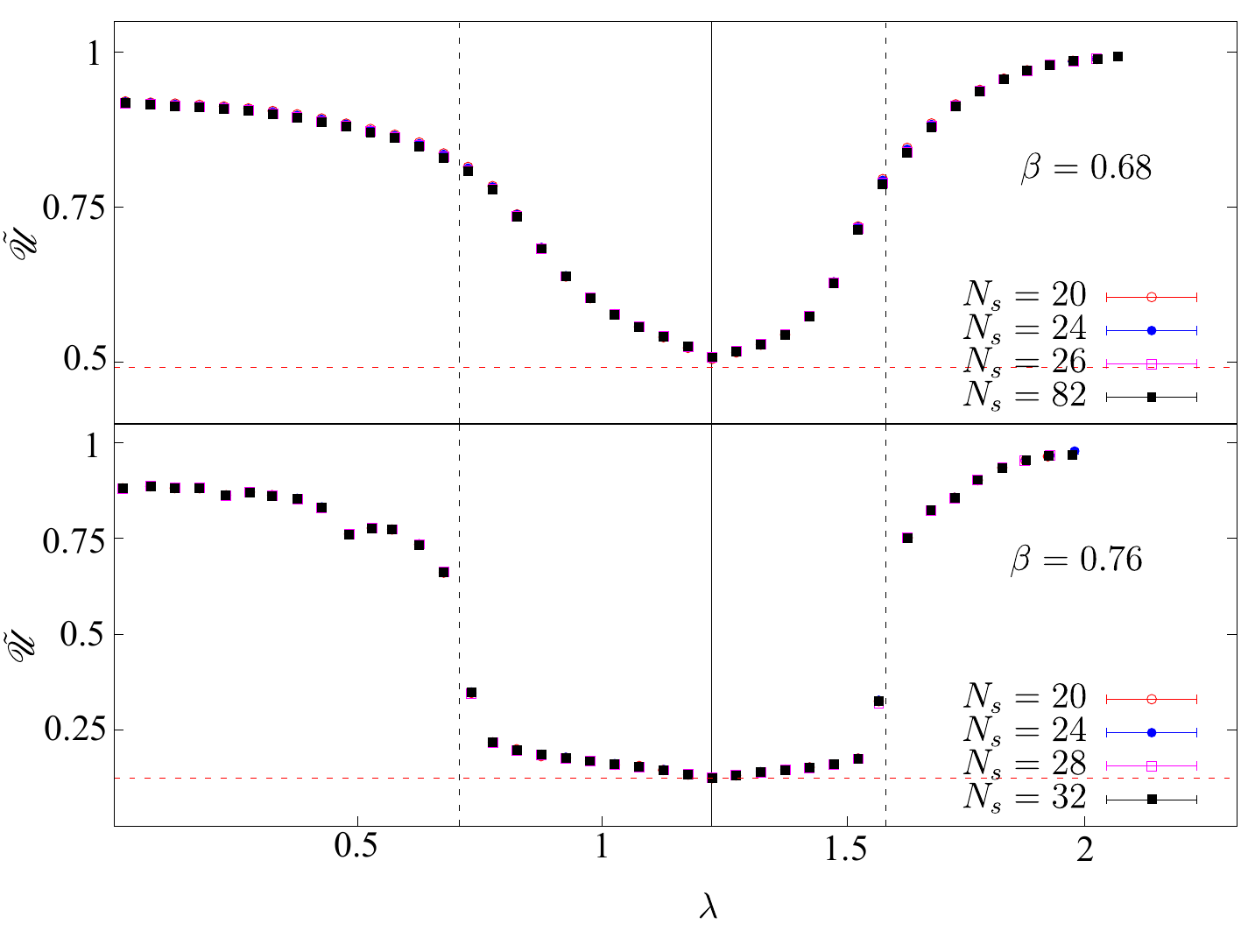}
        \caption{Average number of negative plaquettes touched by a mode (left panel, confined phase (top) and deconfined phase (bottom)). Part of mode touched by a negative plaquette (right panel, confined phase (top) and deconfined phase (bottom)). Horizontal dashed lines correspond to $6(1-\langle U_{\mu \nu} \rangle)$ (left panel)  and $2(1-\langle U_{\mu \nu} \rangle)$ (right panel). }
        \label{fig:negative}
\end{figure}
 In Fig. \ref{fig:negative} we show $\mathscr{U}$ and $\Tilde{\mathscr{U}}$ in the confined and deconfined phase. Low and high modes prefer to live close to clusters of negative plaquettes, with no significant difference between confined and deconfined phase. However, as one crosses the temperature of the deconfinement transition, the number of negative plaquettes decreases significantly, while $\mathscr{U}$ and $\Tilde{\mathscr{U}}$ do not change much. For low modes it is only possible if they become localised near negative plaquettes in the deconfined phase.    
\section{Conclusions}
The finite-temperature deconfinement transition leads to the appearance of localised Dirac modes at the low end of the spectrum in a variety of gauge
theories. 
In Ref. \cite{PhysRevD.104.054513} we have studied the localisation properties of the eigenmodes of the
staggered Dirac operator in the background of $\mathbb{Z}_2$ gauge field configurations on the lattice in 2+1 dimensions. This is the simplest gauge theory displaying a deconfining transition
at finite temperature, and so provides the most basic test of the sea/islands picture of
localisation. This picture is confirmed by this work. A novel result is that the very high modes, near the upper
end of the spectrum, are localised in both phases of the theory. We have also shown that there is a strong correlation between negative plaquettes and localised modes.

\end{document}